\documentclass[pra,usletter,superscriptaddress,twocolumn,amsmath]{revtex4}
\usepackage{amsmath,latexsym,amssymb,amsfonts,graphicx}

\newcommand{\bra}[1]{\mbox{$\langle #1|$}}
\newcommand{\ket}[1]{\mbox{$|#1\rangle$}}

\begin{document}

\title{Choice of Measurement Sets in Qubit Tomography} \author{ Mark D
  de Burgh} \affiliation{School of Physical Sciences, University of
  Queensland, St Lucia, Queensland 4072, Australia} \author{ Nathan K.
  Langford} \affiliation{School of Physical Sciences, University of
  Queensland, St Lucia, Queensland 4072, Australia} \author{ Andrew C.
  Doherty} \affiliation{School of Physical Sciences, University of
  Queensland, St Lucia, Queensland 4072, Australia} \author{ Alexei
  Gilchrist} \affiliation{School of Physical Sciences, University of
  Queensland, St Lucia, Queensland 4072, Australia}

\date{\today}
\begin{abstract}
  Optimal generalized measurements for state estimation are well
  understood.  However, practical quantum state tomography is typically
  performed using a fixed set of projective measurements and the
  question of how to choose these measurements has been largely
  unexplored in the literature. In this work we develop 
  theoretical asymptotic bounds for the average fidelity of pure qubit
  tomography using measurement sets whose axes correspond to vertices
  of Platonic solids. We also present complete simulations of maximum
  likelihood tomography for mixed qubit states using the Platonic solid
  measurements. We show that overcomplete measurement sets can be used
  to improve the accuracy of tomographic reconstructions.
\end{abstract}

 \maketitle

\section{ Introduction and Background }
Quantum tomography \cite{89vogel2847}, the practical estimation of quantum states
through the measurement of large numbers of copies, is of fundamental
importance in the study of quantum mechanics.  With the emergence of quantum
information science, the tomographic reconstruction of finite dimensional
systems \cite{95leonhardt4101} has also become an essential technology for
characterizing the experimental performance of practical quantum gates and
state preparation. Examples include tomography of the polarization states of
light \cite{WhiteAG1999a, LangfordNK2004a} and electronic states of trapped
ions \cite{Haeffner2005, 06reichle838}. These experiments would benefit 
from a systematic study of the optimal
measurement and state estimation strategy to use.  

There has been much
theoretical work in this area, and optimal bounds on state estimation,
and constructions for measurements that achieve these bounds are known
\cite{82holevo, Massar1995, 98derka1571, Vidal1999, Latorre1998}. These bounds require all
copies of the state to be collected and a combined measurement
performed across all the copies. While these collective measurements
are known to be more powerful than performing measurements on each
copy one at a time ({\em local measurements}) \cite{Massar1995}, they
are currently totally impractical.

Qubit tomography is currently done using fixed local projective
measurements \cite{James2001, Paris2004}.  The important question of
which fixed local projective measurement sets to use remains an open
problem, although there are several relevant theoretical
discussions~\cite{Jones1991,Bagan2005a,Gill2002, Bagan2006,
  Rehacek2004, Roy2007}. In practice, measurement sets with the
minimal number of measurements such as those described in
\cite{James2001} have become popular, particularly in optical
experiments (but see \cite{LangfordNK2004a}). This situation arises
since in optical experiments a significant amount of time can be spent
on changing the measurement settings. However, such choices are made
without much quantitative understanding of how the performance of
the tomographic reconstruction is affected by restricting the measurement
sets in this way.

\begin{figure}
\begin{center}
  \includegraphics[width=.45\textwidth]{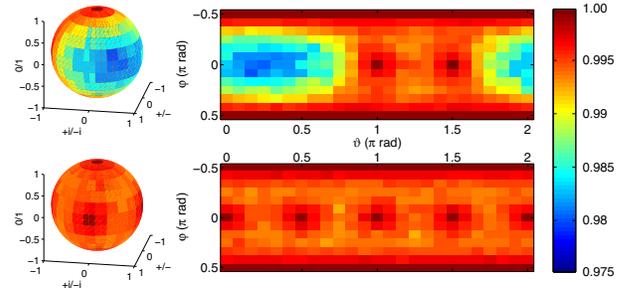}
\end{center}
\caption{Performance of tomographically reconstructed states on the Bloch
  sphere for two measurement sets.  The top set is a popular set of
  measurements (James4) used in the literature \cite{James2001}, the
  bottom set is a more isotropic set composed of measurements along
  the three spatial axes (a cube measurement set), see text for
  details. For each target state the average fidelity with 400
  reconstructed states is plotted by color. The ensemble of
  reconstructed states was generated by adding Posssonian noise to a
  simulated experiment with a total of 4000 counts, and then
  performing maximum-likelihood tomography.}
\label{Fig:tomotheorypaperfig} 
\end{figure}

In this paper we investigate how choice of measurements affects 
the quality of tomographic reconstruction. We follow
Jones~\cite{Jones1991} and investigate a class of measurement sets
based on Platonic solids that provide excellent performance for
tomography using fixed local projective measurements. This class of
measurement sets allows one to trade off performance against the number
of measurement settings, with more overcomplete measurement sets
performing better. For example in Fig.~\ref{Fig:tomotheorypaperfig} we
show that both the average and worst case fidelities of the
tomographic reconstruction are greatly improved by replacing the
popular minimal measurement set of \cite{James2001} with the
overcomplete cube Platonic measurement set.



Platonic solid measurements were first proposed for tomography in
\cite{Jones1991} for the special case of pure states and using mutual
information as a figure of merit. However, this figure of merit makes
it hard to compare these results to more recent work which typically
uses the average fidelity.  We derive new asymptotic bounds for the
performance of Platonic solid measurements for pure states using
average fidelity as a figure of merit. We then present full numerical
simulations of both one and two-qubit tomography performance as a
function of the number of copies of the system, using these
measurement sets.  This provides a detailed study of how information
is acquired during a tomography experiment. To our knowledge such a
study has not been performed in experimental work but would determine
whether tomographic reconstructions are limited by the expected
statistics or by other experimental imperfections, such as drifting
sources.

The average fidelity is the most commonly used
measure of the performance of state reconstruction 
consequently we make use of this figure of merit. 
Much of the original interest in fidelity arises
from the fact that it bounds the distinguishability of quantum states,
see \cite{Fuchs1996}. Recently a true quantum Chernoff bound on the
distinguishability of states has become available \cite{Audenaert2006}
and this quantity is a tighter bound than the fidelity. As
a result we use this quantity as figure of merit also, although the
results do not depend greatly on this choice.

Very recently, Roy and Scott \cite{Roy2007} identified a class of
measurements, including all the Platonic solid measurements, that give
the optimal performance for another figure of merit, the mean squared
Hilbert-Schmidt distance.  Nevertheless, it will turn out that when
average fidelity is used as the figure of merit, there is a range of
performance within the class, with the higher order Platonic solids
performing better. This indicates that the choice of figure of merit
can certainly have a qualitative effect on the comparison between
tomographic procedures.

The structure of the paper is as follows: We will introduce the
tomography problem, and discuss figures of merit for tomography
schemes in section~\ref{sec:qubittomo}. In section~\ref{sec:platonic}
we define the Platonic solid measurements.  In
section~\ref{sec:ComparisonMeasurementSets} we describe some
other measurement sets that have been studied in the literature for
comparison with the Platonic solid measurements.  We
present theoretical pure state
asymptotic bounds on the average fidelity of the tomographic
reconstruction for the different platonic solid measurements in
section~\ref{sec:PureStateTheory}. We then give some intuition for
mixed state tomography in section~\ref{sec:MixedStateTheory}.  After
presenting details of our mixed state simulations for atomic and
photonic qubit systems in section~\ref{sec:simulation} we give our
results for one and two qubit systems in section~\ref{sec:results}.
Finally we will investigate the effects of using different figures of
merit in sections \ref{sec:quantumchernoff} and
\ref{sec:hilbertschmidt}.

\section{ Qubit Tomography }
\label{sec:qubittomo}
A $d$ dimensional quantum state $\rho$ is represented by a
$d{\times}d$ positive semidefinite density matrix, with trace one.
There are $d^2{-}1$ real parameters to be estimated. In the case of optical
experiments such as \cite{WhiteAG1999a,
  LangfordNK2004a} the flux must also be estimated giving $d^2$
real parameters to estimate.

Any setting $l$ of an experimental apparatus designed to measure the
quantum state may be described by a positive operator value measure (POVM). 
Each of the $k$ outcomes of
a measurement setting is represented by a positive semidefinite
operator $O_{lk}$. The operators satisfy $\sum_k O_{lk} = I$. The
probability of observing outcome $k$ is given by the Born rule $p_{lk}
= \mathrm{tr}(\rho\, O_{lk})$.  In state tomography each of the $l$
measurements are performed on a large number of copies and estimates
obtained for each of the $p_{lk}$. We denote these estimates
$\hat{p}_{lk}$. When the number of linearly independent $O_{lk}$
equals or exceeds the number of parameters to estimate, our
measurements are known as informationally complete \cite{Caves2002}.
If these probability estimates were perfect, so that
$\hat{p}_{lk}=p_{lk}$, it would be possible to
reconstruct the state exactly. In this case there is a set of
operators $R_{lk}$, known as a dual basis or dual frame, 
 such that $\rho = \sum_{lk} p_{lk} R_{lk}$. (When the number of
 linearly independent $O_{lk}$ exceeds the number of parameters to be
 estimated the dual frame is not unique.) Since our
probability estimates are not perfect due to the finite sample size,
they may well be inconsistent with any such reconstruction. In this
case it is common practise to resort to a maximum likelihood estimate
of the state \cite{James2001, Paris2004}.  We look at maximum
likelihood reconstruction in detail in section
\ref{Sec:MaximumLikelihood}

To judge the quality of reconstruction we need a measure of similarity
between the true state $\rho$ and the reconstructed state
$\hat{\rho}$.  In this paper we will concentrate on two measures. The
first is the commonly used \emph{fidelity}:
\begin{eqnarray}
\label{eqn:fidelity}
f(\rho, \hat{\rho}) &=& \left[\mathrm{tr} \sqrt{\sqrt{\hat{\rho}} 
\rho \sqrt{\hat{\rho}}}\right]^2.
\end{eqnarray}
where $0\le f(\rho, \hat{\rho})\le1$ and $f=1$ implies
$\rho=\hat{\rho}$.  The second is the quantity that determines the
recently derived \emph{quantum
  Chernoff bound} \cite{Audenaert2006}:
\begin{equation}
\label{eqn:quantumchernoffbound}
\lambda_{cb}(\rho, \hat{\rho}) = \underset{0\le s \le 1}{\mathrm{min}} \mathrm{tr}[\rho^s \hat{\rho}^{1-s}].    
\end{equation}
where as for the fidelity,  $0\le \lambda_{cb}(\rho,
\hat{\rho})\le1$ and $\lambda_{cb}=1$ also implies $\rho =
\hat{\rho}$. 

The quantum Chernoff bound has a clear physical interpretation that can
be understood by considering the following situation.  Suppose an
experimentalist wishes to determine if her state preparation device
creates the state $\rho$ or the state
$\sigma$. Time is limited and she only has $N$ identically prepared
copies of the 
state to work with. She is assumed to be able to perform any measurement,
including a collective measurement, in her efforts to distinguish the
two states. The result of \cite{Audenaert2006} is that the probability
of her making an error is asymptotically
$\mathrm{P_e} \approx e^{N \ln \lambda_{cb}(\rho, \sigma) }$ where
$\lambda_{cb}$ is called the quantum Chernoff bound. When one of
the states is pure, the quantum Chernoff bound coincides with the 
fidelity. 
Also the square root of the fidelity is always an 
upper bound on the quantum Chernoff bound. \cite{Fuchs1996,
  Audenaert2006}. Somewhat confusingly this quantity has also
been termed the fidelity in other works and serves as an alternative
 measure of distinguishability. We have chosen the definition in 
 (\ref{eqn:fidelity}) for it's appealing interpretation when one
 of the states is pure, and because all of the theoretical results
 we mention are derived using this definition.
  As the quantum Chernoff bound has a clear
  physical interpretation even when both states are mixed, we
consider it a better motivated measure of the distinguishability of $\rho$ and $\sigma$.

Next we define quantities which rate the performance of a tomography
scheme for a particular true state $\rho$, using the fidelity and
quantum Chernoff bound.  The figure of merit we adopt is based on how
distinguishable the true state is from the reconstructed state, when
averaged over measurement outcomes and an appropriate ensemble of true
states. (An interesting alternative approach was advocated by
Blume-Kohout and Hayden~\cite{Blumekohout2006}.) A tomography experiment produces a list of
outcomes of measurements performed on $N$ copies of the true state
$\rho$.  These outcomes are used to obtain an estimate of the state
$\hat{\rho}$, and the fidelity (or quantum Chernoff bound) between the
true and estimated states is calculated. Averaging over 
the outcomes of many such tomography experiments we obtain the 
\emph{point fidelity}
\begin{equation}
F(\rho, N) = \sum_{\hat{\rho}} p_\rho(\hat{\rho})f(\rho, \hat{\rho})
\end{equation}
where $p_\rho(\hat{\rho})$ is the probability of estimating state
$\hat{\rho}$ given the true state $\rho$. Or \emph{point Chernoff
  bound}:
\begin{equation}
\lambda_{pt}(\rho, N) = \sum_{\hat{\rho}} p_\rho(\hat{\rho})\lambda_{cb}(\rho, \hat{\rho}).
\end{equation}
Finally averaging the point fidelity (or point Chernoff bound) over
all possible true states, we obtain the \emph{average fidelity}:
\begin{eqnarray}
\label{eqn:averagefidelity}
F = \int\!\! d\rho \;F(\rho, N).
\end{eqnarray}
or the \emph{average Chernoff bound}:
\begin{eqnarray}
\label{eqn:averagechernoff}
\lambda_{av} = \int\!\! d\rho \;\lambda_{pt}(\rho, N).
\end{eqnarray}
These are the figures of merit we will use to rate a tomography
scheme.

To perform the averages over true states in
equations~(\ref{eqn:averagefidelity}) and~(\ref{eqn:averagechernoff}),
we must choose a prior probability distribution over true states. For
pure states there is one clear choice --- the Haar measure.  For mixed
states the choice is not so clear, see for example~\cite{zyczkowski2001}. We will
make use of two priors, one
based on the fidelity, and the other on the quantum Chernoff bound.

The first, known as the Bures prior, is the density of states induced
by defining a volume element using the distance $1/2\arccos f(\rho,
\rho + d\rho)$.  It is a natural choice when using the fidelity as our
measure of similarity between states. It has also been argued that the
Bures prior corresponds to maximal randomness of the input states
\cite{Hall1998}. The Bures prior is unitarily invariant, and it gives a
radial probability density on the Bloch sphere:
\begin{equation}
p_B(r) = \frac{4}{\pi}\frac{r^2}{\sqrt{1-r^2}}
\end{equation}

\begin{figure}
\begin{center}
  \includegraphics[width=.45\textwidth]{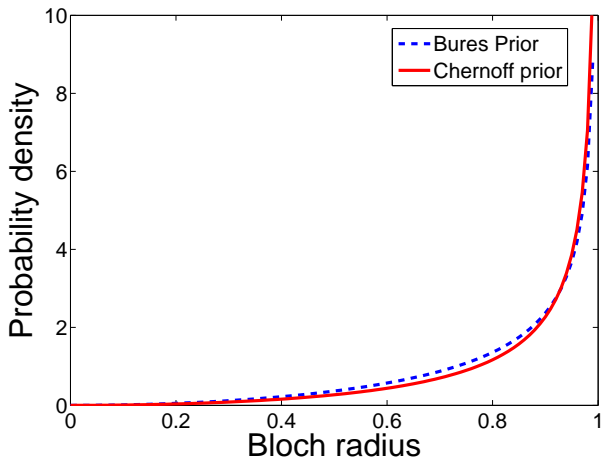}
\end{center}
\caption{Plot of probability density against Bloch radius for the Bures
  and Chernoff priors over states. Both distributions have densities
  which increase with state purity. The Chernoff prior is slightly
  more skewed towards the pure states than the Bures prior.}
\label{Fig:PriorsVsRadius} 
\end{figure}

When the quantum Chernoff bound is used as the measure of similarity,
it is natural to use a prior over states based on this.  The natural
metric induced by the quantum Chernoff bound was derived in
\cite{Audenaert2006}. Some of the authors of \cite{Audenaert2006} have
also derived the corresponding volume
element in the space of states which can be used as a prior on mixed states
\cite{MunozTapia2007}.  In Appendix \ref{sec:chenoffevals} we present
our general derivation for states of arbitrary dimension. The
resulting Chernoff prior is unitarily invariant and the probability
density over
eigenvalues is given by:
\begin{equation}
p(\lambda_1..\lambda_M) = \frac{C}{\sqrt{\lambda_1..\lambda_M}}
\prod_{k<j} \frac{(\lambda_k - \lambda_j)^2}{(\sqrt{\lambda_j} + \sqrt{\lambda_k})^2}
\end{equation}
where the eigenvalues are also constrained to satisfy $\sum_{j=1}^M
\lambda_j = 1$, and C is a normalisation constant. For single qubit
states this corresponds to a radial probability density on the Bloch
sphere given by:
\begin{equation}
p_{CB}(r) = \frac{2}{\pi-2}\frac{1-\sqrt{1-r^2}}{\sqrt{1-r^2}}. 
\end{equation}
A comparison of the density of the Bures and Chernoff priors is given in
Fig.~\ref{Fig:PriorsVsRadius}.  Both these priors have increasing
density with state purity.  The Chernoff prior is slightly more skewed
towards pure states, than the Bures prior.

We will present all of our mixed state results in terms of average fidelity
with states drawn according to the Bures prior first, as this has
become the popular way of
assessing mixed state tomography. We will then present our detailed
results using the average Chernoff bound figure of merit with states
drawn according to the 
Chernoff prior. We believe that this figure of merit is better
motivated and thus it is important check that our qualitative
conclusions do not
depend on this choice of figure of merit.

In other works the mean square Hilbert-Schmidt distance and 
the mean squared Euclidean distance between Bloch vectors have been used
as a figures of merit for tomography \cite{Scott2006,
  Rehacek2004}. For single qubits these measures are equivalent
  up to constant factors. The Hilbert-Schmidt distance is defined as
$D_{\mathrm{hs}}(\rho, \sigma)
= \sqrt{\mathrm{tr}(A^\dag A)}$ ,where $A = \rho - \sigma$.  It is
again related to the probability
of error when an experimentalist must determine if she has a state
$\rho$ or state $\sigma$, however in this case the experimentalist has
only a single copy to measure.  Under the assumption that both states
are equally likely before the measurement, the probability of error
in distinguishing single qubit states $\rho$ and $\sigma$ is 
$P_e=\frac{1}{2} - \frac{1}{2\sqrt{2}} D_{\mathrm{hs}}(\rho, \sigma)$. 
In order to get an analytically
tractable measure of distinguishability \cite{Scott2006,
  Rehacek2004} square $D_{\mathrm{hs}}$ before averaging
this quantity over measurement outcomes and states as we have
done for fidelity and the quantum Chernoff bound.
This figure of merit makes it possible to find the
optimal measurement sets analytically but after squaring and averaging
this quantity is not as well motivated.  We will examine
the effect of using this figure of merit in
section~\ref{sec:hilbertschmidt} and show that it can lead to
qualitatively different behaviour.

   
\section{ Platonic solid measurements }
\label{sec:platonic}

The state density matrix of a qubit can be written as:
\begin{equation}
\label{eq:DensityMatrix}
\rho = \frac{I + \vec{r}.\vec{\sigma}}{2}
\end{equation}
where $\vec{r}$ is a real three dimensional vector known as the Bloch
vector, and $\vec{\sigma}$ is a vector of the Pauli matrices
$\sigma_x$, $\sigma_y$ and $\sigma_z$. In the special case of pure
states, the Bloch vector is unit length. A two-outcome local
projective measurement on the $l$th measurement setting can be
represented by two projectors:
\begin{equation}
\label{eq:ProjectiveMeasurement}
O(\pm \vec{m_l}) = \frac{I \pm \vec{m_l}.\vec{\sigma}}{2}
\end{equation}
where $\vec{m_l}$ is a real three dimensional unit Bloch vector.  The
orthogonal outcomes correspond to opposite directions on the Bloch
sphere.

As states are isotropically arranged in the Bloch ball, one
intuitively expects tomography performance to depend on how
isotropically spaced our measurement vectors are on the Bloch sphere.
This intuition was supported (and made precise) by the asymptotic pure
state results of \cite{Jones1991}, and is demonstrated in
Fig.~\ref{Fig:tomotheorypaperfig}.  In this paper we will concentrate
on the measurement sets whose Bloch vectors form Platonic solids.
These are the five convex regular polyhedra: tetrahedron, cube,
octahedron, dodecahedron and icosahedron. Such shapes are highly
isotropic and thus would be expected to give excellent tomographic
performance.

Mathematically, the idea of isotropically spacing points on a sphere
is also captured by the concept of a \emph{spherical $t$-design}. A
description of $t$-designs is beyond the scope of this paper but note 
that the properties of $t$-designs mean that they are excellent sets
of points for performing
numerical estimation of
integrals over the sphere with the quality of the approximation being
expected to improve as $t$ is increased. This suggests that good
measurement sets for tomography may well be associated with $t$-designs.  
In fact, popular measurement sets for tomography of general
quantum systems include the so-called
symmetric informationally complete POVMs \emph{SicPOVMs} \cite{Renes2004} 
which are spherical 2-designs. We note that the tetrahedron
measurement set described below is the
SicPOVM for a single qubit, and all the Platonic solids are also
2-designs. In fact it is well known \cite{Hardin1996} that 
the cube and octagon are 3-designs and the
dodecahedron and icosahedron are 5-designs.  (Note that a $t$-design is also a
$t^\prime$-design for all $t^\prime{<}t$).  This further motivates us
to concentrate on the performance of the Platonic solids.

There are two possible ways of defining Platonic measurement sets. We
define a \emph{Platonic solid measurement set} to be a set of $L$
different measurement settings whose $2L$ Bloch vectors $\pm \vec{m_1}
\dots \pm \vec{m_L}$ match the $2L$ centres of the faces of a Platonic
solid. We use this definition so that an octagon has eight measurement
directions. An alternative definition would be to define the Platonic
solid measurement as the set of Bloch vectors matching the vertices of
a Platonic solid. This definition would result in the same set of
shapes because the centres of the faces of one Platonic solid, match
the vertices of another Platonic solid. Such solids are said to be
duals of each other. The dual pairs are the tetrahedron with itself,
the cube with the octagon and the dodecahedron with the icosahedron. We find
the first definition more appealing and use it throughout the paper.

The experiments we will be modelling are described in detail in
section \ref{sec:simulation}.  In the atomic qubit experiment, and the
dual-detector photonic experiment each measurement has two orthogonal
outcomes. These correspond to opposite directions on the Bloch sphere.
Thus two opposite directions are measured simultaneously by a single
measurement. All the Platonic solids except the tetrahedron have a
face opposite every face and thus these Platonic solid measurements
can be measured in this way while the tetrahedron cannot.  (Attempting
to measure
the tetrahedron directions would result in us measuring the octagon
instead). For the single-detector photonic qubit experimental model,
we will require $L$ measurement settings to obtain the $L$ outcomes,
as the orthogonal outcome is not detected. This enables a tetrahedron
measurement to be meaningfully measured in this case. Direct
measurement of the tetrahedron is possible in principle and schemes
have been proposed~\cite{Decker2003,
  Englert2005}, that would however require significantly
greater experimental resources so we will not consider this
possibility in the following.

We will also be interested in two qubit states. The experimentally
simplest measurements are projective measurements performed locally on
both subsystems (recording joint outcome probabilities). While we
would expect measurement sets that include measurements entangled
across both subsystems to give improved results, these measurements
are difficult to perform. We will examine the performance of Platonic
solid measurements on each subsystem. That is measurements of the
form:
\begin{equation}
\label{Eq:AtomicTensors}
O(\pm \vec{m_i}) \otimes O(\pm \vec{m_j})
\end{equation}
where both $i$ and $j$ range over the $L$ measurements of the same
Platonic solid.

\section{ Measurement sets for comparison }
\label{sec:ComparisonMeasurementSets}


For simulations of one and two qubit optical experiments (single-detector
configuration) we will compare our results to the popular measurement sets
described in \cite{James2001}. The one qubit state measurement set
consists of measuring polarizations \{ H, V, D, R \}, that is the
horizontal and vertical linear polarizations, the diagonal linear
polarization [$\ket{D}{=}(\ket{H}{+}\ket{V})/\sqrt{2}$], and the right circular polarization 
[$\ket{R}{=}(\ket{H}{+}i\ket{V})/\sqrt{2}$]. We refer to this
measurement set as James4.  The two qubit state measurement
set is \{
HH, HV, VV, VH, RH, RV, DV, DH, DR, DD, RD, HD, VD, VL, HL, RL \}
where for example the measurement setting HL means measuring horizontal polarization on the first qubit
subsystem and left circular polarization [$\ket{L}{=}(\ket{H}{-}i\ket{V})/\sqrt{2}$] on the second qubit
subsystems \footnote{The reason for measuring L instead of R seems purely a 
historical accident. A measurement set consisting of the tensor product of two James4 sets
would equally suffice. Nevertheless we will maintain the L in the James16 measurement set in this paper.}.  
We refer to this measurement set as James16. (Notice
that this measurement set does not form complete POVMs. 
The implementation we have in mind is explained fully in Section \ref{sec:simulation})

In the case of two qubits we also consider a modified version of the SicPOVM of
\cite{Renes2004} for comparison. We call this measurement a projective
SicPOVM. The original
SicPOVM for two qubit states, is an entangled measurement in which a
single measurement produces one of 16 outcomes. Experimentally this
measurement is difficult to realize. Our projective SicPOVM is a
measurement set of 16 settings, in which each setting projects onto
one of the directions of the SicPOVM. While still entangled this
measurement should be simpler to implement experimentally than the
SicPOVM.

\section{ Pure state estimation of qubits with Platonic solids}
\label{sec:PureStateTheory}

The bound on average fidelity for collective measurements is known to
be: $F{\le}(N{+}1)/(N{+}2) \approx 1{-}1/N$ where the approximation is
good for large $N$.  In fact this bound is achievable
with collective measurements \cite{Massar1995}.  When restricted to
local measurements, measuring the three
orthogonal axes of the Bloch 
sphere (a cube measurement) is known to give an average fidelity
bounded by $F{\le}1{-}13/(12N)$ and again this bound is achievable
\cite{Bagan2002}.  We have derived average fidelity bounds for the
other Platonic solids. The derivation only slightly generalizes
Appendix~B of \cite{Bagan2005a}.  We give an overview of our
derivation here, leaving the details to Appendix~\ref{app:cramerrao}.

These bounds on average tomography performance are a consequence of
the \emph{Cram\'{e}r-Rao bound} of classical statistics. The Cram\'{e}r-Rao
bound concerns the problem of sampling from a probability
distribution, where the probability distribution is determined by some
parameters.  The goal is to estimate these parameters based on the
results of sampling the distribution. The Cram\'{e}r-Rao bound states that
the variance of an unbiassed estimator is asymptotically lower bounded
by $O(1/N)$, and the coefficient of this scaling is given by the
\emph{Fisher information} (the definition may be found in
Appendix~\ref{app:cramerrao}).

In a tomography experiment, with a fixed set of measurements, the
outcome probability distribution is fixed by the parameters of the
quantum state we are trying to estimate, and we can directly apply the
Cram\'{e}r-Rao bound. The procedure is as follows:

\begin{enumerate}
\item Consider a tomography experiment on a quantum state with
  parameters $\eta$, and its fidelity to the state estimated from the
  tomographic reconstruction with parameters $\hat{\eta}$.  In an
  asymptotic regime our estimates will be close to the true state, so
  we can expand the fidelity in a Taylor series about the true state.
  
\item Calculating the expected value of the fidelity over the
  different measurement outcomes, we obtain an asymptotic expression
  for the point fidelity which is a function of the covariance matrix
  of the estimated parameters.
  
\item By applying the Cram\'{e}r-Rao bound to the covariance matrix of the
  estimated parameters, we can obtain an upper bound on the point
  fidelity which is a function only of the number of each kind of
  measurement we have made, and the state parameters.
  
\item Finally by averaging this point fidelity expression uniformly
  over the state parameters (using the Haar measure) we obtain the
  average fidelity bound.
\end{enumerate}

While our expressions are analytic up to the final step, we use
numerical integration to average over the state parameters using the
Haar measure and obtain our final asymptotic results.
  
We find $F{\le}1{-}y/N$ where $y$ is 1.083 for the cube (in agreement
with the value of 13/12 in \cite{Bagan2002}), 1.049 for the octahedron, 1.018
for the dodecahedron, and 1.008 for the icosahedron. It is well known
in classical
statistics that the Cram\'{e}r-Rao bound can be achieved using a maximum
likelihood estimator. Hence the above bounds are also tight.  These
results show that for pure states the performance of the Platonic
solids measurements in tomography is close to the 
collective measurement bound of $y{=}1$.

Recall that the quantum Chernoff bound is equal to the fidelity if one
of the states involved is pure. When, as here, both states are pure
the HilbertSchmidt distance is also an equivalent measure as, 
$D_{hs}^2{=}(1{-}f)/2$. So that the choice of figure of merit does 
not affect the results of this section at all.

\section { Mixed state qubit tomography}
\label{sec:MixedStateTheory}

When the state to be reconstructed can be mixed an analytical
treatment is more complicated since the requirement that the density
matrix be positive semidefinite leads to constraints on the allowed
set of parameters. The situation is also complicated by the way in
which the fidelity function depends on the mixedness of the state to
be reconstructed. We partially address both these issues in this
section, with the main aim to develop intuition. For quantitative
results we will resort to simulation.

For mixed state qubit tomography, the state Bloch vector is no longer unit
length. The state vector is now:
\begin{equation} 
\label{Eqn:MixedStateVector}
\vec{r} = [r_x, r_y, r_z]^T 
\end{equation}
where $r_x{=}r\cos(\phi)\sin(\theta)$,
$r_y{=}r\sin(\phi)\sin(\theta)$, $r_z{=}r\cos(\theta)$ and $r$
satisfies $r{\le}1$. This constraint on the length of the Bloch vector
reflects the requirement that the density matrix is positive semidefinite.

\begin{figure}
\begin{center}
  \includegraphics[width=3.25in]{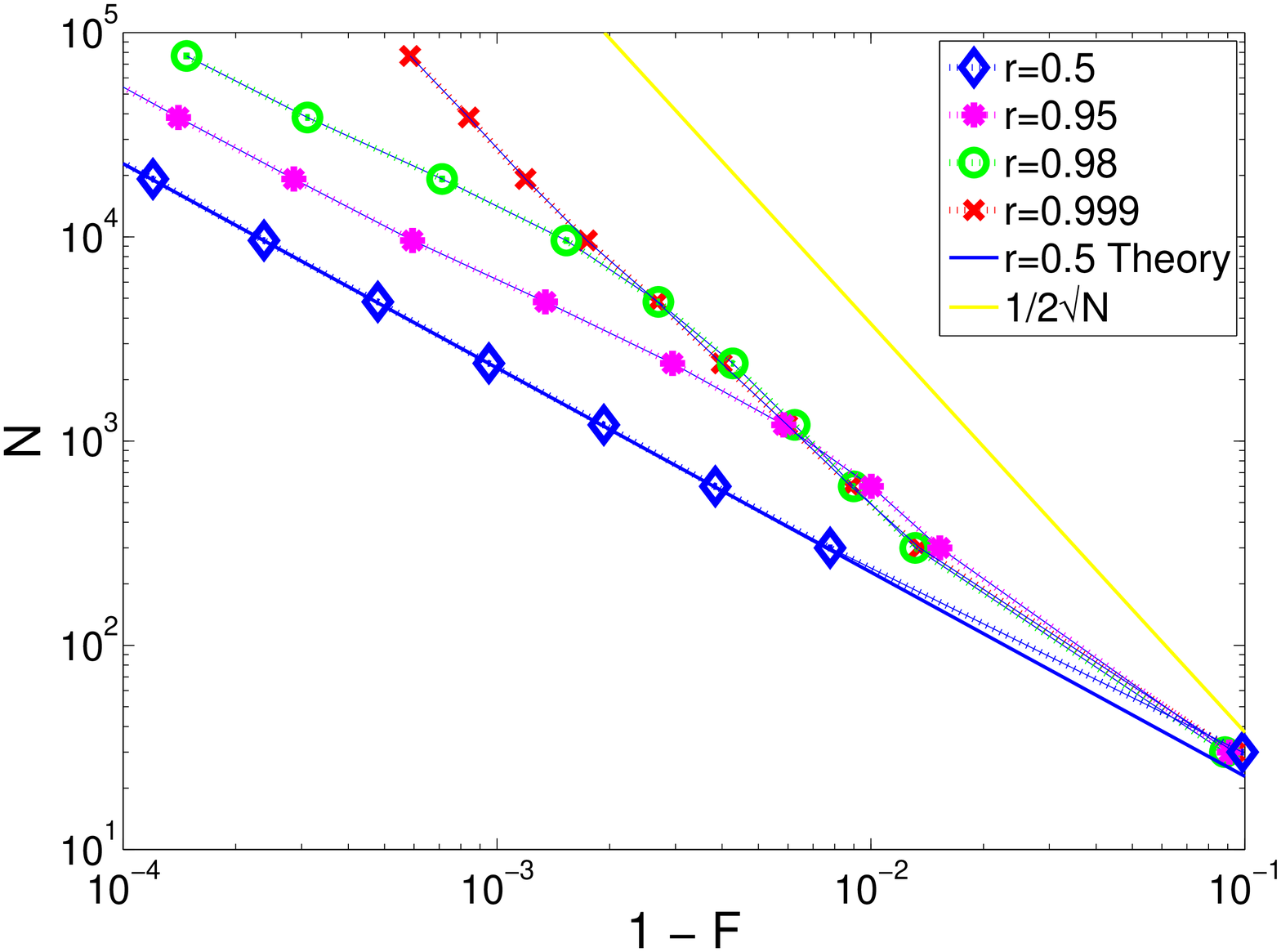}
\end{center}
\caption{Numerical simulations of atomic point fidelities for a state 
  rotated 45 degrees about the x-axis from the +ve z-axis are plotted
  against radius on the Bloch sphere. The cube measurement set was
  used for these simulations. Observe mixed states always have a $1/N$
  scaling. States that are close to pure start (after we get to a
  sufficiently large $N$ to obtain a sensible state estimate) with a
  $1/\sqrt{N}$ dependence and transition to $1/N$ scaling with larger $N$. This
  knee occurs at larger $N$ for purer states. Also the statistical
  theory of equation (\ref{eqn:MixedAverageCubeFidelity}) is plotted
  for $r{=}0.5$ giving excellent agreement}
\label{Fig:AtomicOffAxisPointFidelityVsRadius} 
\end{figure}

Mixed state tomography introduces two new complications which we
discuss in this section. Both concern behaviour close to the boundary
of the Bloch sphere. The first is due solely to the definition of the 
mixed state fidelity function:
\begin{equation}
f(\vec{r}, \vec{\hat{r}}) = \frac{1 + \vec{r}.\vec{\hat{r}} + 
\sqrt{1-r^2}\sqrt{1-\hat{r}^2}}{2}.
\end{equation}
where $\vec{\hat{r}}$ is the estimated state vector. It can be
understood by considering only errors in the Bloch vector length $r$
\cite{Bagan2006}. Suppose $r{=}1{-}\delta$ and our estimate is
$\hat{r}{=}r{-}\epsilon$. Then provided $\epsilon{\ll}\delta$ and both
are small we find: $F{\approx}1{-}\epsilon^2 / 8$.
From the statistical arguments of the previous section we would expect
$\epsilon$ to scale asymptotically like $1/\sqrt{N}$, so we expect to
see all point fidelities (except for perfectly pure states) to scale
like $F{\approx}1{-}1/8N$ once $N$ is large enough that our errors are
well inside the Bloch ball. Alternatively if $\delta{\ll}\epsilon$, we
find that the point fidelities scale like $F{\approx}1{-}\epsilon /
2$. Thus the general behaviour of point fidelities is a transition from
$1/\sqrt{N}$ scaling at low $N$ to a scaling of $1/N$ at large $N$.
The transition takes place when the error associated with the state
reconstruction $\epsilon$ becomes small enough that $\hat{r}+\epsilon$
lies inside the Bloch ball. Thus the location of the transition happens at
higher $N$ for higher purity
states.  This is illustrated in
Fig.~\ref{Fig:AtomicOffAxisPointFidelityVsRadius} where numerical
simulations of point fidelities for states with different radial
parameters are shown.
The Bures prior has a radial distribution
that is strongly peaked at $r=1$ and it turns out that when point
fidelities are averaged over this prior the $1/\sqrt{N}$ dependence of
the point fidelity
persists long enough that the dependence of the average fidelity on
$N$ is not $N^{-1}$ but rather $N^{-3/4}$ \cite{Bagan2004}.

The second complication arises since our estimated physical state has
to be positive semidefinite.  For qubits this is equivalent to
$\hat{r}{\le}1$. An estimator that predicts states lying outside the
boundary of the Bloch ball can be improved by mapping these states
onto the boundary of the Bloch ball which always moves the estimate closer to
the true state. Maximum likelihood estimation is a specific example of
an estimator that does exactly this. This procedure creates a biassed estimator
that nevertheless performs better than any unbiassed one. The na\"{i}ve
Cram\'{e}r-Rao estimation is therefore not appropriate.

For states well away from the boundary, we can ignore this positivity
constraint, and use the unbiased statistical methods of the previous
section extended to three parameters to calculate the cube measurement
point fidelity:
\begin{equation}
\label{eqn:MixedAverageCubeFidelity}
F(\vec{r}, N) \le 1-\frac{3}{4N}\frac{3-3r^2+2 ( r_y^2 r_x^2+ r_z^2 r_x^2+ r_z^2 r_y^2)}{1- r^2}.
\end{equation}
This value is included in
Fig.~\ref{Fig:AtomicOffAxisPointFidelityVsRadius} for the case when
$r{=}0.5$ and found to give excellent agreement with the numerical
simulations.

\section{Simulation details}
\label{sec:simulation}
The main result of this paper is the numerical simulation of
tomography using Platonic solid measurements. The simulations are
performed in the following way.
 
\begin{enumerate}
\item Choose $N_s$ states $\rho_i$ at random according to the Bures
  (or Chernoff) prior. For our simulations $N_s$ was chosen to be
  10,000.
\item For each $\rho_i$, simulate measurements on $N_c$ copies of the
  state (divided equally between the measurement sets) generating
  $N_c$ measurement outcomes $\chi_i^j$. For our simulations we chose
  $N_c{=}100{,}000$.
\item Perform maximum-likelihood state estimation based on increasing
  numbers $N$ of these $\chi_i^1..\chi_i^N$ obtaining state estimates
  $\hat{\rho}_i(N)$.
\item Calculate the fidelity $f(\hat{\rho}_i(N), \rho_i)$ [or Chernoff
  bound $\lambda_{cb}(\hat{\rho}_i(N), \rho_i)$] between each state
  estimate and true state.
\item Calculate the average fidelity $F(N){=}\sum_{i=1}^{N_s}
  f(\hat{\rho}_i(N), \rho_i)/N_s$ [or the average Chernoff bound:
  $\lambda_{av}(N){=}\sum_{i=1}^{N_s} \lambda_{cb}(\hat{\rho}_i(N),
  \rho_i)/N_s$].
\end{enumerate}

The Monte-Carlo averaging over states and measurement outcomes is
performed simultaneously. 

We present our results on a log-log plot of
number of copies required against $1{-}F(N)$. To find the number of
copies required to achieve a target average fidelity, one can select
the target average fidelity on the horizontal axis, and tracing
vertically upward, find the number of copies required to achieve that
target average fidelity. The inserts on the graphs are enlargements of
the largest $N$ section of the graph.  The width of the lines on these
inserts are approximately equal to the one standard deviation 
statistical error in average fidelity due to our choice of $N_s$.

We simulate two main scenarios, tomography on atomic qubits, and
tomography on photonic qubits.

\subsection{Atomic Qubits}

The kind of experiment we model here is an atomic qubit as in
experiments on trapped ions, such as~\cite{Haeffner2005, 06reichle838}
represented by
two metastable energy levels $\ket{0}$ and $\ket{1}$, and an auxiliary
level $\ket{r}$ used for performing the measurement with a laser
driving the transition
$\ket{0}{\rightarrow}\ket{r}$ leading to fluorescence if the state was
initially in $|0\rangle$
\cite{Cirac2004}. Arbitrary single qubit projective measurements are
made available by a first laser pulse tuned to the
$\ket{0}{\rightarrow}\ket{1}$ transition followed by turning a laser
on the $\ket{0}{\rightarrow}\ket{r}$ transition and looking for this
fluorescence. We assume the dominant source of error is statistical
fluctuation due to the finite number of copies of the state, and
neglect all other sources of error. Many of these could be taken
into account by replacing our projective measurements with suitable
POVMs. However the effects such as drifts in either state preparation
or measurement settings would necessitate a different approach.

In this limit these single qubit system measurements are well described by Eqns.
(\ref{eq:DensityMatrix}), (\ref{eq:ProjectiveMeasurement}) and
counts for each outcome $k$ of a measurement setting $l$ 
drawn from a multinomial distribution:
\begin{equation}
\label{eqn:multinomial}
p(n_{l1}..n_{lK}) = \frac{N_l}{n_{l1}..n_{lK}} p_{l1}^{n_{l1}} p_{l1}^{n_{l2}}...p_{lK}^{n_{lK}}
\end{equation}
where $N_l$ copies of the system are measured with measurement setting
$l$, $p_{lk}{=}\mathrm{tr}(\rho O_{lk})$ and $K$ is number of possible
outcomes for the measurement (2 for qubit systems and 4 for two-qubit systems).

For two-qubit systems, the measurement operators $O_{lk}$ take the form of
Eq.~(\ref{Eq:AtomicTensors}) and now the index $l$ ranges over all 
combinations of single qubit measurement settings on the subsystems. 
Experimentally this corresponds to noting the joint probability of the 
four possible outcomes of fluorescence and no fluorescence in each subsystem,
for all combinations of measurement settings on the subsystems.
 
\subsection{Photonic qubits}

\begin{figure}
\begin{center}
  \includegraphics[width=3.25in]{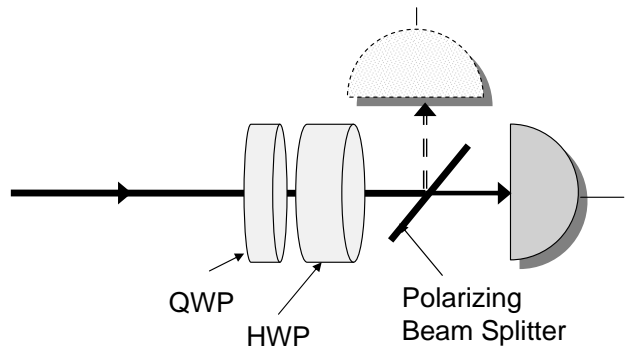}
\end{center}
\caption{The single photonic qubit experimental configuration consists of a
quarter wave plate and a half wave plate to set the basis, and a polarizing
beam splitter followed by a photon detector to analyse in that basis. We call
this a single-detector configuration. An additional detector can be placed on
the reflected port in a dual-detector configuration, in which case the
statistics of the counts can be easily reduced to the atomic case by waiting
until a fixed number of counts are obtained for each basis setting.}
\label{Fig:expfig} 
\end{figure}

Our model for experiments on single photonic qubits is shown in
Fig.~\ref{Fig:expfig}.  In this model our qubits are represented by
the polarization state of photons issuing from a source with
Poissonian arrival times. There are two configurations to consider:
a single-detector configuration and a dual-detector
configuration.

In the first configuration, the source is directed through a quarter
wave plate, a half wave plate and polarizing beam splitter (or
vertical polarizer) and finally a single photon counter
\cite{James2001}. The angles of the fast axes of both of the wave
plates can be rotated, allowing the projection onto vertical
polarization fixed by the final polarizer to be rotated into any
polarization state that the experimenter requires. Photon counts are accumulated
over a time $t$ for each measurement setting $m_l$. The total photons
incident on the measurement apparatus during this time is drawn from 
the Poisson distribution:
\begin{equation}
\label{eqn:poisson}
p(N_l) = \frac{e^{-\mathcal{N}_f t} (\mathcal{N}_f t)^{N_l}}{N_l!}
\end{equation}
where $\mathcal{N}_f$ is the mean photon flux.
The number of these photons reaching the detector $n_l$ is then drawn 
from a binomial distribution with probability $p_l = \mathrm{tr}(\rho O_l)$.  
The flux is estimated from a set of $r$ measurements satisfying
\begin{equation}
\label{Eq:fluxEstRequirement}
\sum_{i=1}^{r} O_i = I
\end{equation}
via
\begin{equation}
\mathcal{N}_f = \sum_{i=1}^r n_i / t.
\end{equation}

For the Platonic solid measurements, the four measurements of
the tetrahedron satisfy (\ref{Eq:fluxEstRequirement}) and for higher
order Platonic solids each axis has two measurements satisfying
(\ref{Eq:fluxEstRequirement}). In the later case an average is made
over all the axes to obtain an improved flux estimate.

Adding a detector on the reflected port of the polarizing beam splitter gives
a dual-detector configuration and
potentially better flux estimation. In this configuration both orthogonal
directions are measured simultaneously, and the total flux is estimated by
summing the two detector counts. By waiting until a set number of total counts are
obtained for each measurement set, the statistics of the counts will reduce to the 
atomic case covered above.

We also simulate two photon state tomography. A source produces photon
pairs whose arrivals are again Poisson distributed according (\ref{eqn:poisson}).  
The single photon configuration (single- and dual-detectors) are duplicated (giving two and
four detectors). Now only coincidences are counted between the two
single photon configurations.
The number of co-incidences are drawn from the multinomial distribution of
(\ref{eqn:multinomial}).  For the single-detector configuration for each qubit
we estimate the flux via
\begin{equation}
\mathcal{N}_f = \frac{\sum_{j=1}^r\sum_{i=1}^r n_{ij} }{ r^2 t / 4 }.
\end{equation}

\subsection{Maximum likelihood state reconstruction}
\label{Sec:MaximumLikelihood}

After simulating measurements on all copies of the state we have
$n_{lk}$ detections corresponding to each measurement operator
$O_{lk}$. Maximum likelihood reconstruction is the problem of finding
the state $\rho$ which maximizes the likelihood
$p(...n_{lk}...|\rho)$ or equivalently minimizes $-\log p(...
n_{lk}...|\rho)$. To make our inversions efficiently solvable, we convert
(an approximate version of) this problem into a semidefinite program.

For atomic (single and multiple) qubit systems, and for each measurement 
setting $l$, the outcomes follow the multinomial distribution of
(\ref{eqn:multinomial}).

We will approximate these statistics with a Gaussian distribution 
with mean $\mu_{lk}{=}p_{lk}(\rho) N_l$ and a covariance
matrix $V_l$ with diagonal entries $(V_l)_{ii}{=}N_l p^e_{li} ( 1{-}p^e_{li} )$,
and off-diagonal entries $(V_l)_{ij}{=}-N_l p^e_{li} p^e_{lj}$ 
where $p^e_{lk}{=}n_{lk} / N_l$. Since the outcomes of different 
measurement settings are statistically 
independent, the combined likelihood function is
$p(...n_{lk}...|\rho){=}\prod_l p(n_{l1}..n_{lK})$. 
Defining $\vec{n_l}$ as the vector with $k$th element $n_{lk}$, and 
$\vec{\mu_l}$ as the vector with $k$th element $\mu_{lk}$, our approximate
problem is to find a state $\rho$ which minimises
\begin{equation}
-\log p(...n_{lk}...|\rho) \propto \sum_l 
	(\vec{n_l} - \vec{\mu_l})^T V_l^{-1} (\vec{n_l} - \vec{\mu_l})
\end{equation}
We have assumed that the term in the exponential of the Gaussian dominates
the likelihood and used experimental probabilities $p^e_{lk}$ instead of
$p_{lk}(\rho)$ in the covariance matrices $V_l$ to make the problem
convex. This enables the problem to be converted into a semidefinite 
program and solved with the SeDuMi semidefinite program solver 
software \cite{SeDuMi}.

For the dual-detector simulations of single photon and two photon 
experiments the maximum likelihood reconstruction is 
identical to the reconstruction for the atomic case (if we wait for a 
fixed number of counts). 

For single-detector simulations of single photon and two photon
experiments, the counts are Poisson distributed and
as we only measure one outcome per measurement setting, all outcomes
are statistically independent.
We make a Gaussian approximation with the expected number
of counts for measurement outcome $O_l$ equal to the variance
in those counts i.e. 
$\mu_{l}{=}\sigma_{l}^2{=}\mathcal{N}_{l}{=}p_{l} \mathcal{N}_f t$ 
where $p_l{=}\mathrm{tr}(\rho O_l)$.
Our approximate problem is to find the state $\rho$ which minimise 
the log likelihood
\begin{eqnarray*}
  \label{eq:loglikelihood}
  -\log[ p(n_1,n_2,...|\rho)]&=& \sum_{l} \frac
   {(n_{l}-\mathcal{N}_{l})^2}{2\mathcal{N}_{l}}\\ & =&
   \frac{\mathcal{N}_{f} t}{2}\sum_{l} \frac
   {(p^\mathrm{e}_{l}-p_{l}(\rho))^2}{p_{l}(\rho)}, 
\end{eqnarray*}
where $p^\mathrm{e}_{l} = n_l / (\mathcal{N}_f t)$.
In this case we were able to use $\mathcal{N}_{l}$ for mean and
variance without losing the convexity of the problem.

\section{Results}
\label{sec:results}
\subsection{Single atomic qubit results}

\begin{figure*}[htb]
\begin{tabular}{cc}
(a) 1 atomic qubit & (c) 1 photonic qubit\\
\includegraphics[width=.35\textwidth]{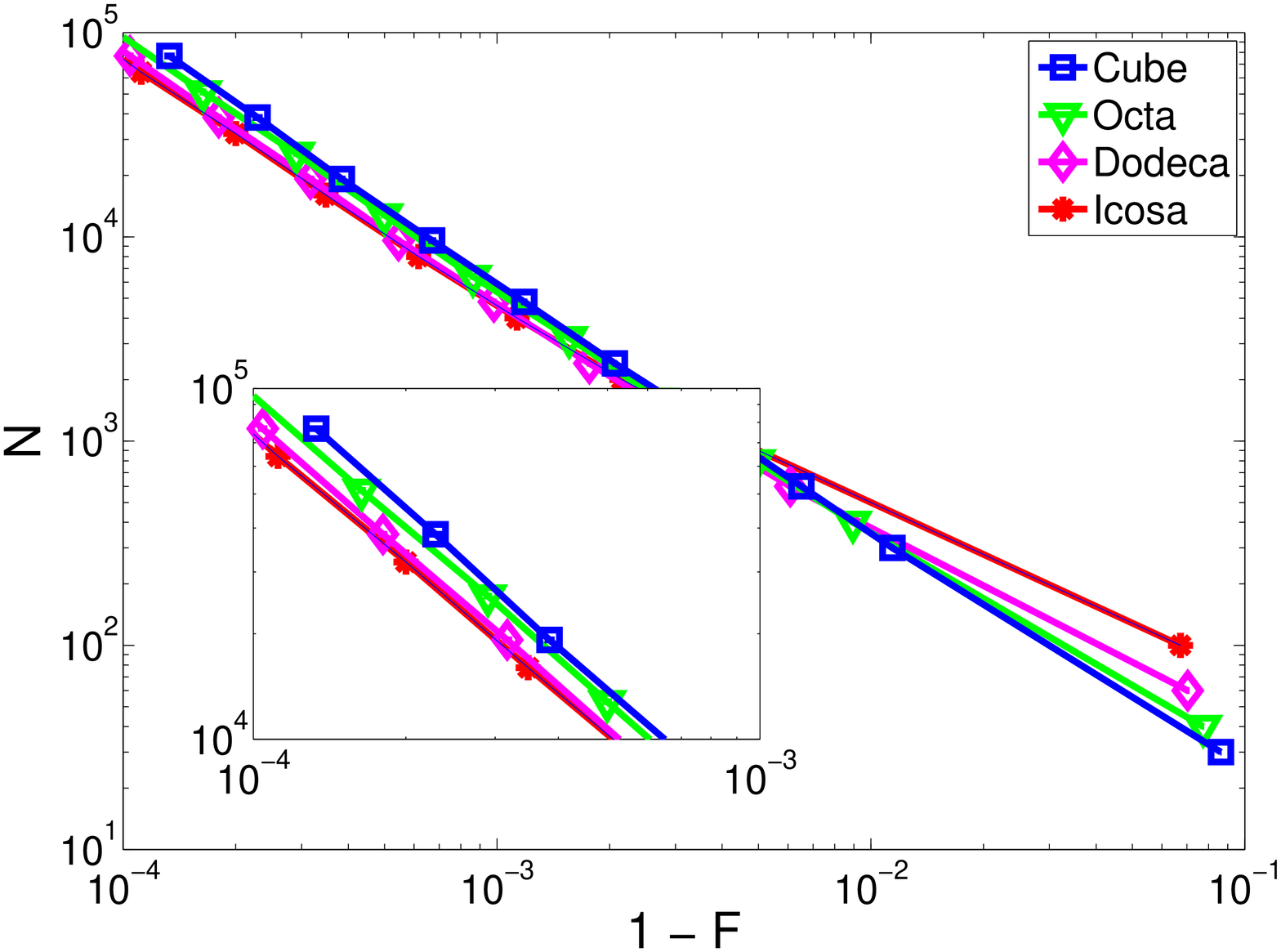}
  &\includegraphics[width=.35\textwidth]{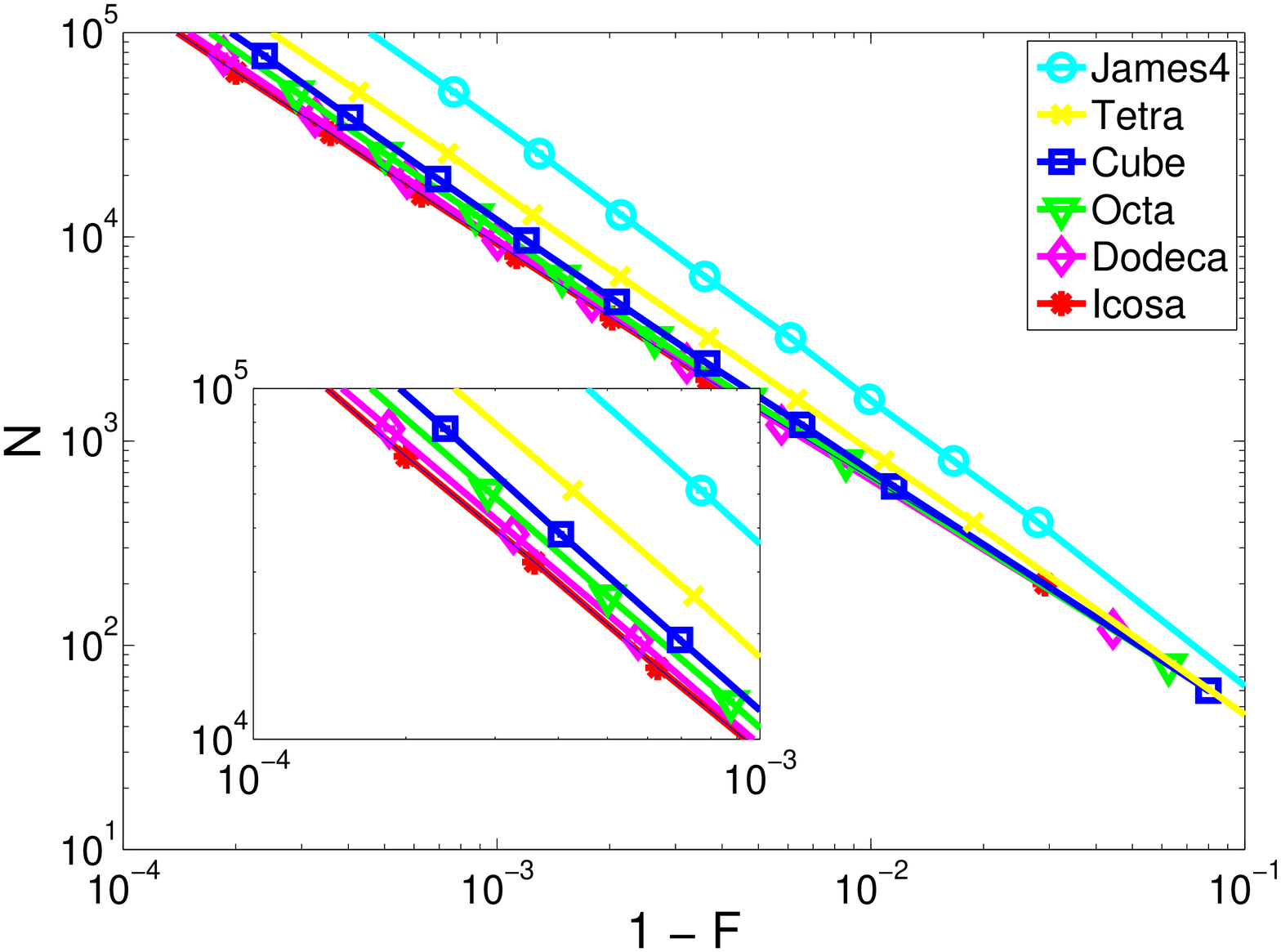}\\
(b) 2 atomic qubits & (d) 2 photonic qubits\\
   \includegraphics[width=.35\textwidth]{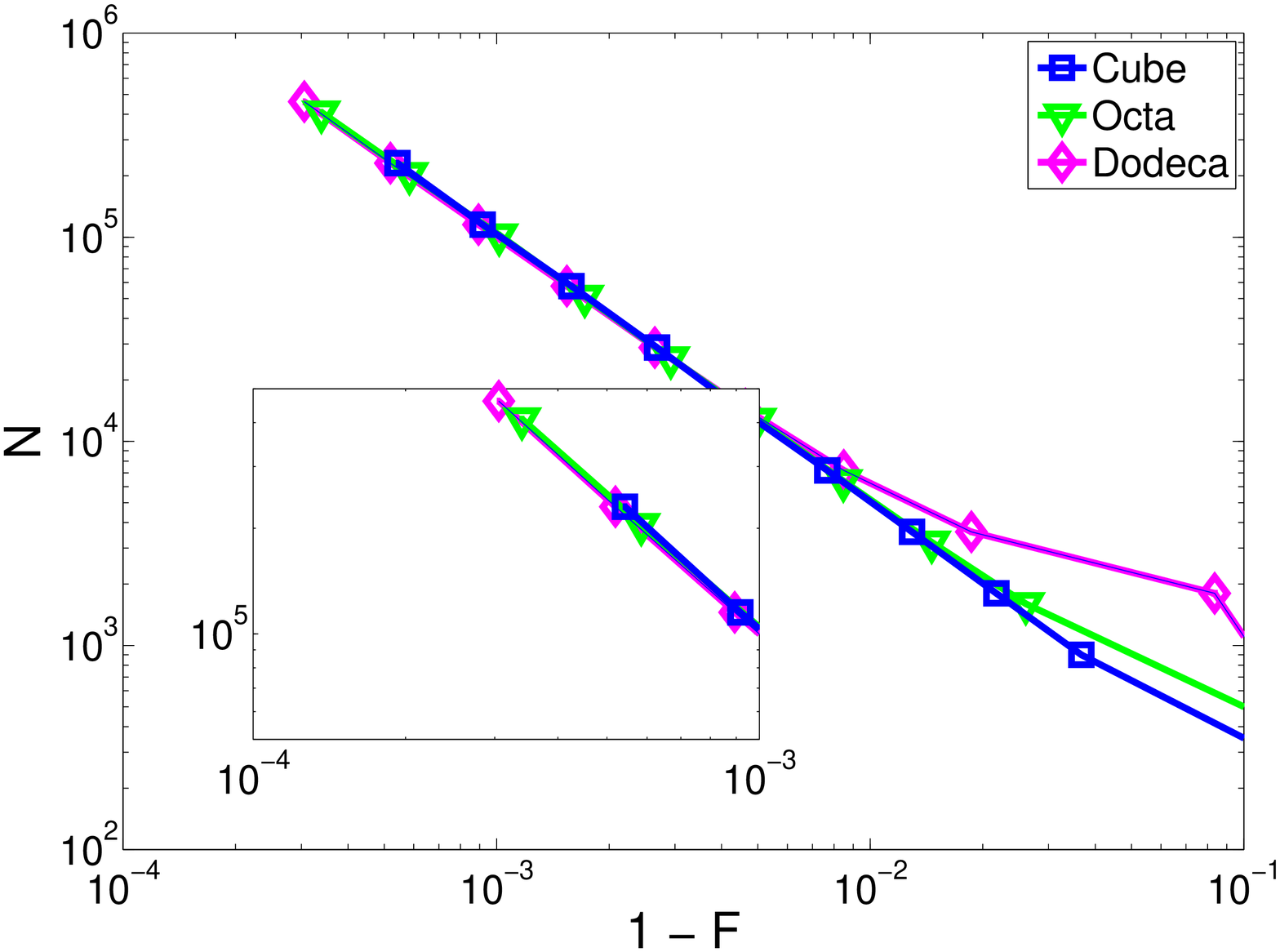}
  &\includegraphics[width=.35\textwidth]{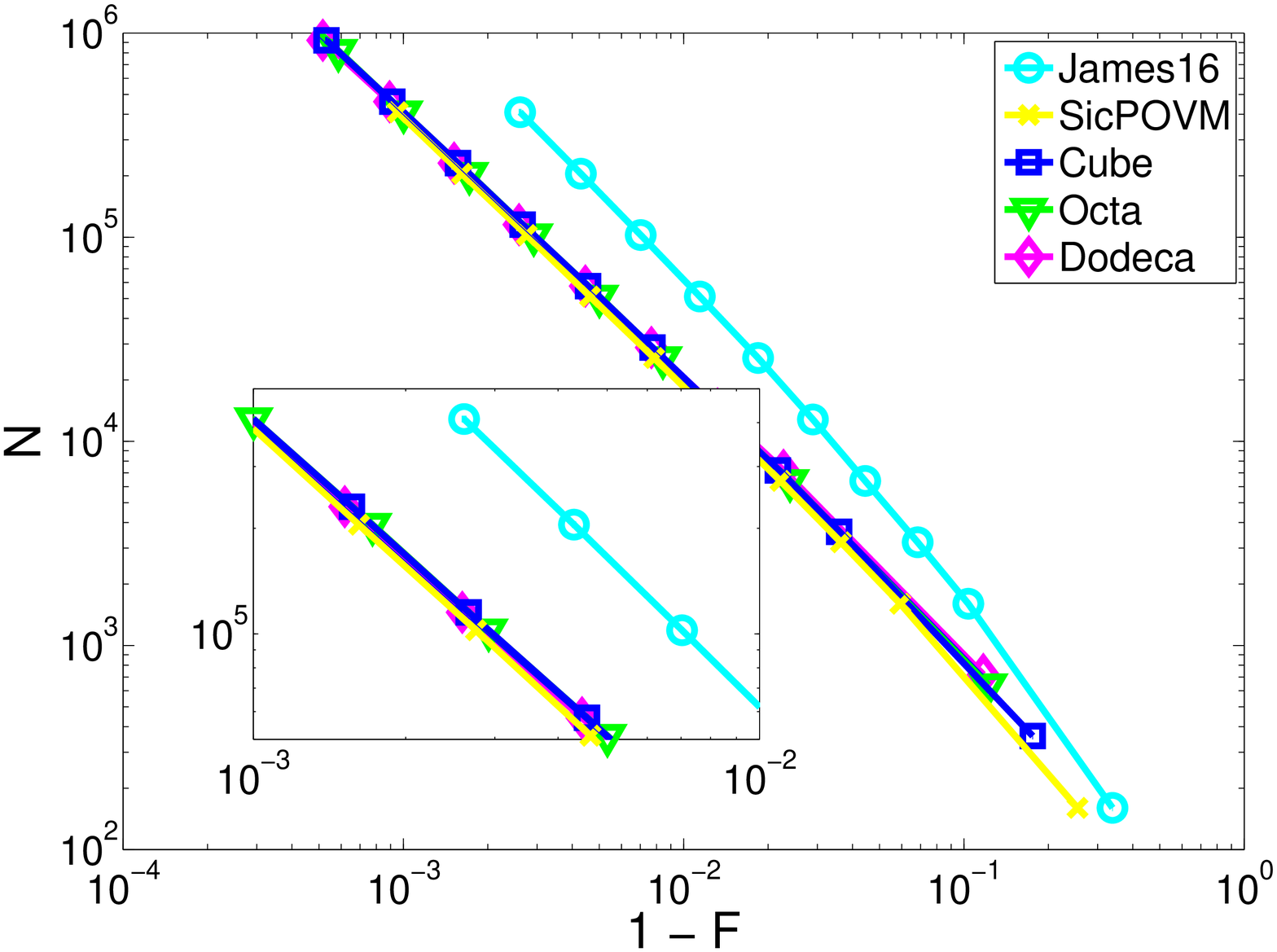}
\end{tabular}
\caption{Tomography simulation results with a Bures prior. The plot is the
number of copies of the state against $1{-}F$ where $F$ is the average
fidelity. To read, select the desired average fidelity on the horizontal axis,
and trace vertically to find the number of copies of the state required to
achieve that fidelity. Each plot shows the performance of various Platonic
solids. Tomography for: 
(a)  single atomic qubit states.  The tetrahedron can not
be measured in this configuration. 
(b) two atomic qubit states.
(c) single photonic qubit with single-detector configuration. 
(d) two photonic qubit with single-detector configuration. 
}
\label{Fig:BuresFid}
\end{figure*}

Simulation results for atomic qubit states are shown in
Fig.~\ref{Fig:BuresFid}(a).  The tetrahedron is not plotted
because for atomic measurements orthogonal outcomes are always measured
simultaneously so, as noted above, the tetrahedron measurement does
not arise without more experimental effort. Observe that, for
sufficiently large $N$, increasing the order of the Platonic solid improves
tomography performance. Recall that $N$ is the {\it total} number of
systems measured so this result holds despite the fact that each
outcome probability is estimated from a smaller number of
measurements. The improvement of the
icosahedron over the dodecahedron is only marginal but the difference
in performance from the other measurements is statistically significant.

\subsection{Two atomic qubit state results}

The simulation results for two qubit atomic states are shown in
Fig.~\ref{Fig:BuresFid}(b).  The performance is effectively
the same for all the Platonic solids and experimentally one should
choose the cube tensor measurement to minimize the time consumed by
changes in measurement settings.


\subsection{Single photonic qubit results}

The single photonic qubit results using a single photon counter are
shown in Fig.~\ref{Fig:BuresFid}(c). In addition to the
Platonic solid measurement sets, we have also included, the popular
measurement sets proposed by James \emph{et al.}\ in \cite{James2001}.
This set is labelled James4.  The most striking aspect of these
results is how poorly the James4 measurement set performs compared to
the Platonic solids.  It is a strong recommendation that
experimentalists using this measurement set should switch to Platonic
solid measurements.  The performance of dual Platonic solids is
similar (the cube and octagon, and the dodecahedron and the
icosahedron), so we recommend using the cube or dodecahedron
measurements.



\begin{figure*}[htb]
\begin{tabular}{cc}
(a) 1 atomic qubit & (c) 1 photonic qubit\\
  \includegraphics[width=.35\textwidth]{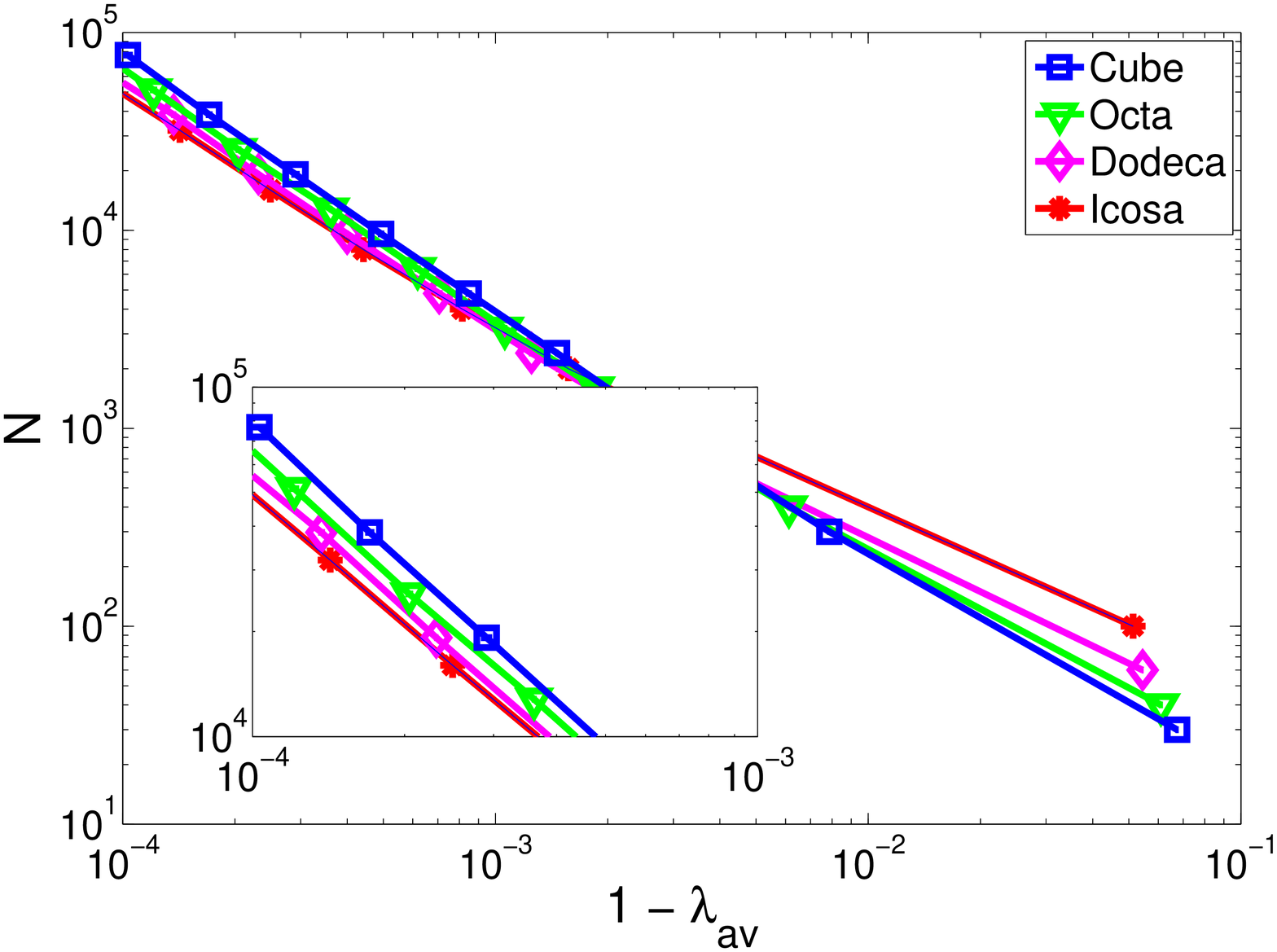}&
  \includegraphics[width=.35\textwidth]{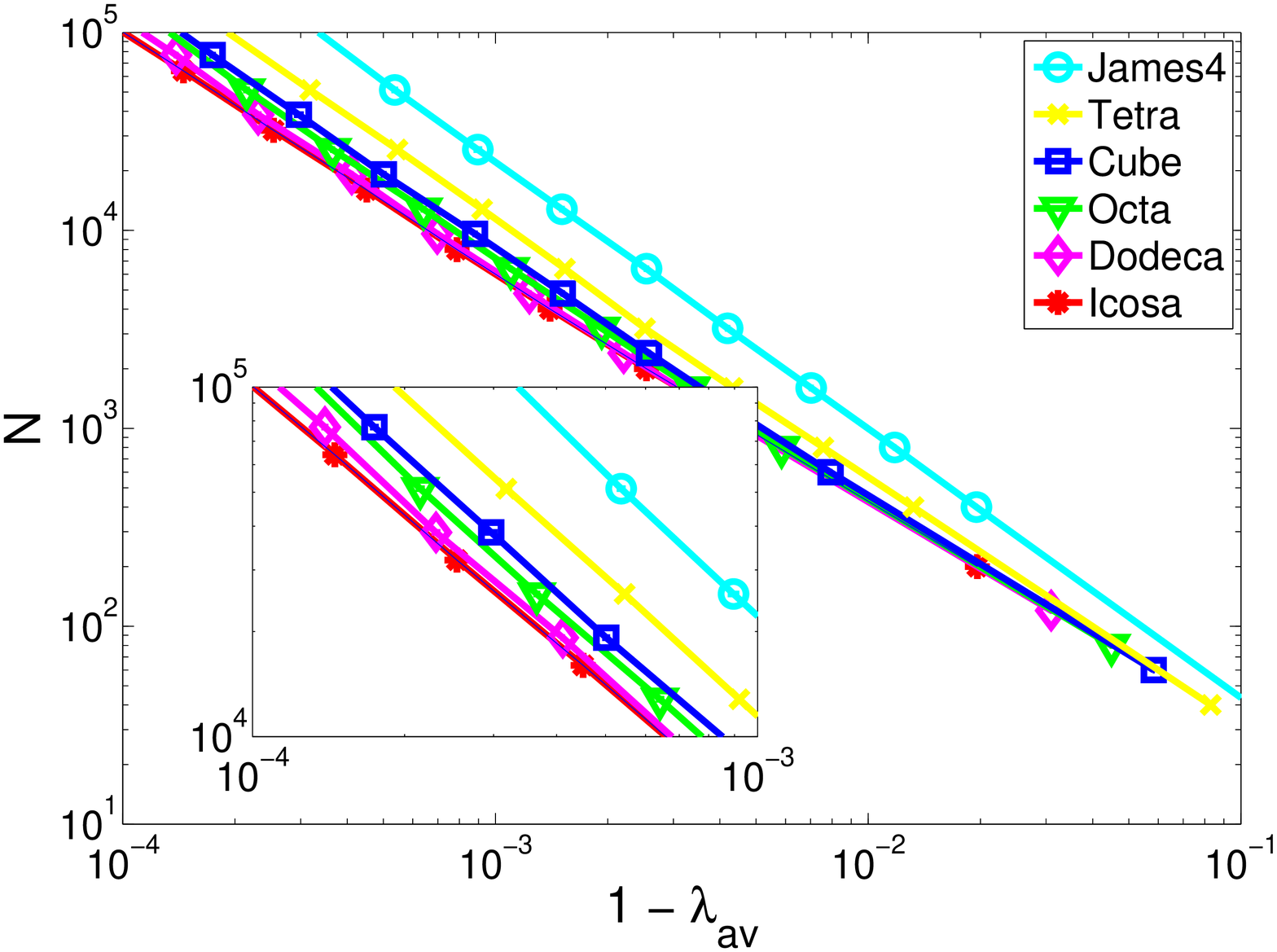}\\
(b) 2 atomic qubits & (d) 2 photonic qubits\\
  \includegraphics[width=.32\textwidth]{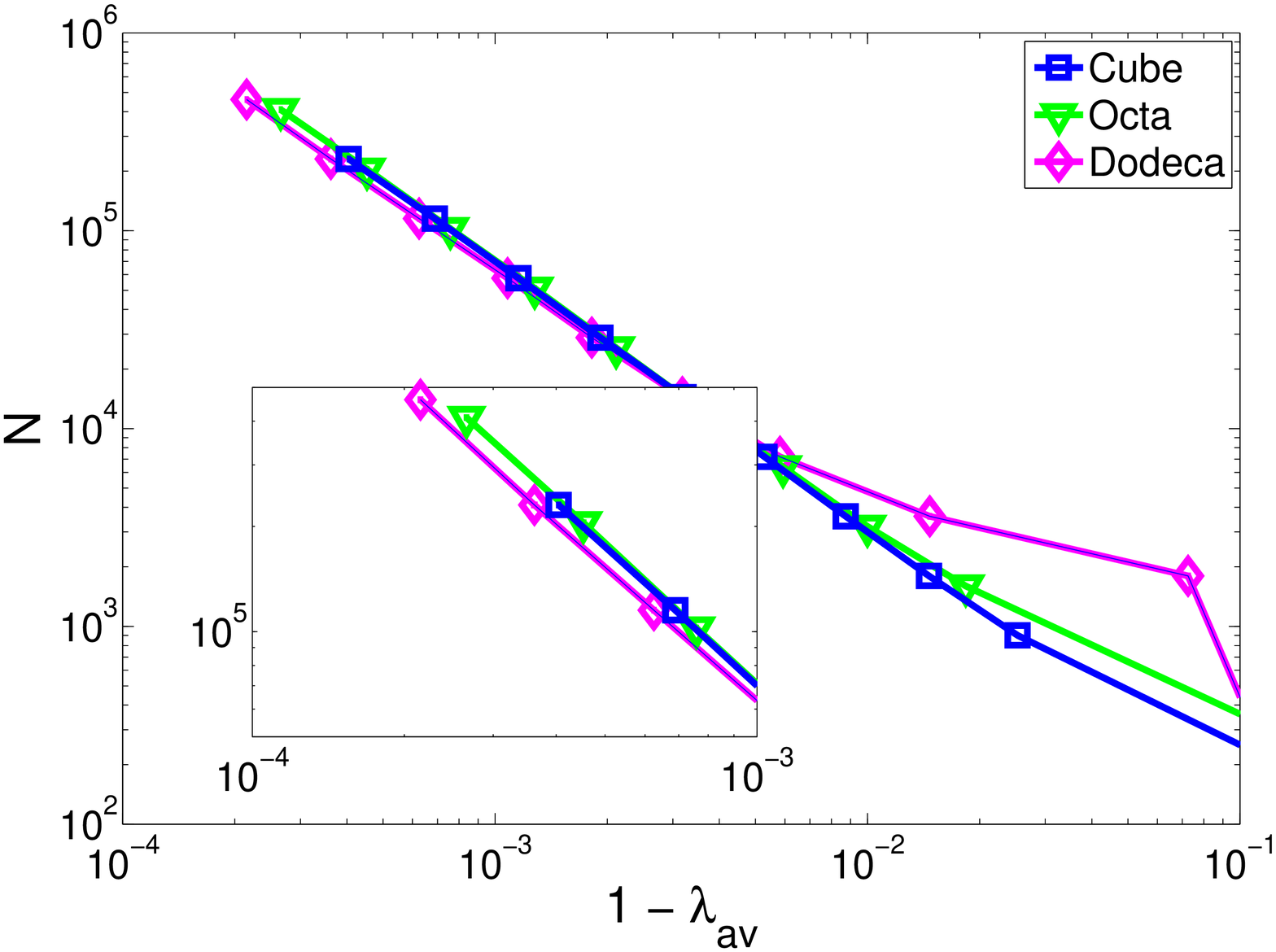}&
  \includegraphics[width=.32\textwidth]{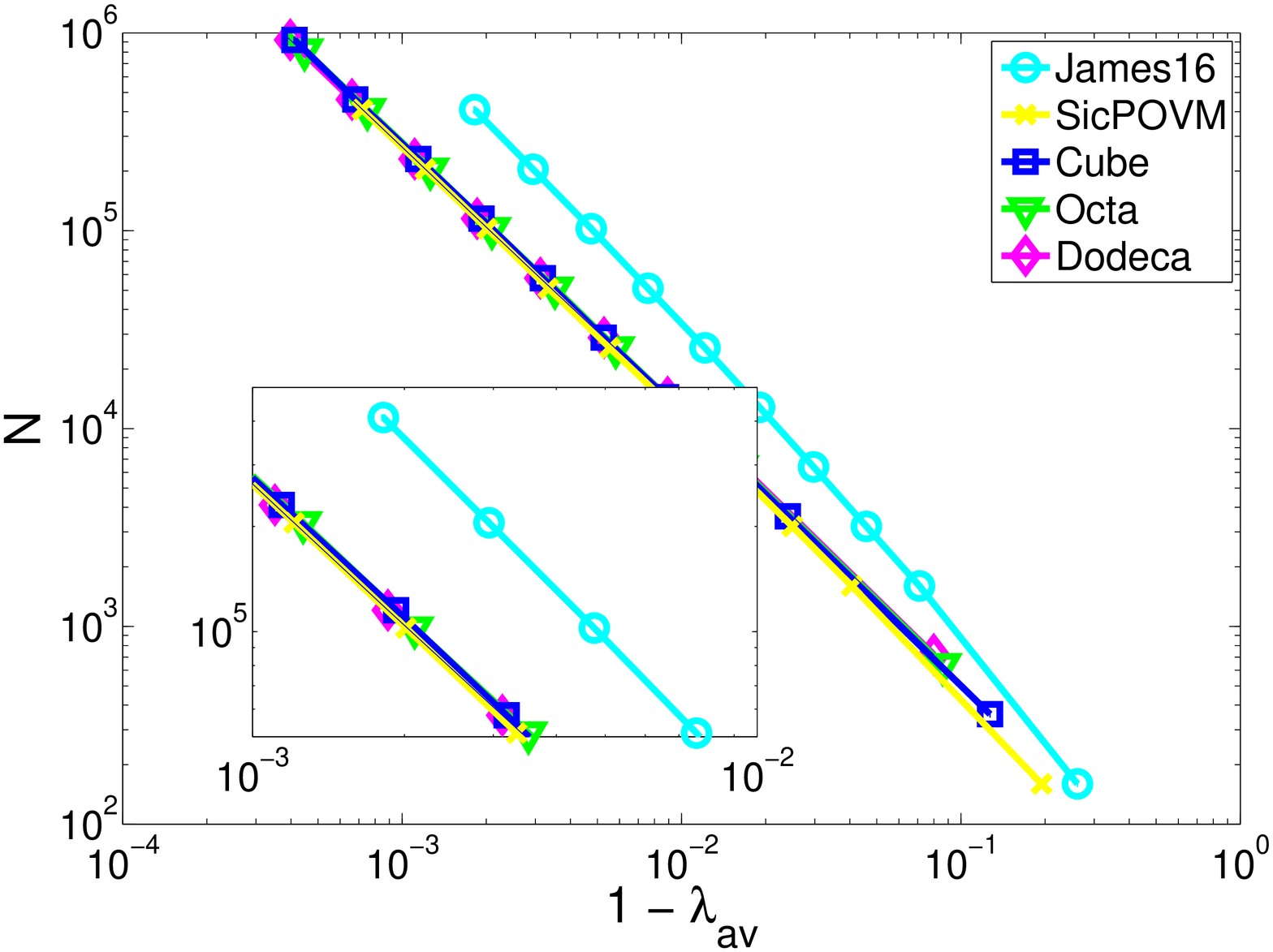}\\
\end{tabular}

\caption{Tomography simulation results with the prior induced by the quantum Chernoff
bound. The number of copies is plotted against $1{-}\lambda_{av}$, where $\lambda_{av}$ is the average
quantum Chernoff bound. Tomography for
(a) a single atomic qubit,
(b) two atomic qubits,
(c) a single photonic qubit with a single-detector configuration,
(d) two photonic qubits using single-detector configurations. 
}
\label{Fig:Chernoff}
\end{figure*}

\subsection{Two photonic qubit state results}

The photonic two-qubit simulation results with 
single-detectors (one for each qubit subsystem) are shown in Fig.~\ref{Fig:BuresFid}(d).  In
addition to the Platonic solid measurement sets, we have also
included, the popular measurement sets proposed by James \emph{et al.}
in \cite{James2001} labelled James16, and the projective SicPOVM
measurements.

Again the most striking result is how poorly the James16 measurement
set performs compared to tensor products of Platonic solids.  It is
our strong recommendation that experimentalists using these settings
change to a measurement consisting of tensor products of cube
measurements or if measurement setting changes consume too much time,
at least to tensors of the tetrahedral measurements.

We also observe that while the projective SicPOVM performed the best,
its improvement was negligible over the cube tensors.  There seems to
be negligible advantage to going to Platonic solids any higher order
than the cube.



\section{Using the Quantum Chernoff bound as a figure of
  merit}
\label{sec:quantumchernoff}

As discussed in section \ref{sec:qubittomo} we find the use of the
average quantum Chernoff bound to be the most
physically appealing figure of merit. It is thus important to see if
our conclusions hold when the quantum Chernoff
bound is used instead of the fidelity.

We repeated our tomography simulations using the average quantum
Chernoff bound as the figure of merit, and the Chernoff bound prior
over states. The results for one and two atomic qubits are shown in
Figs.~\ref{Fig:Chernoff}(a) and~\ref{Fig:Chernoff}(b) respectively.  The results for one and two photonic
qubits with single-detector configurations are shown in
Figs.~\ref{Fig:Chernoff}(c) and~\ref{Fig:Chernoff}(d) respectively.

In all cases the graphs are almost identical to those using a Bures
prior and average fidelity as a figure of merit.  It is clear that all
of our conclusions from the average fidelity are correct, and hold
also when using the quantum Chernoff figure of merit.

\section{Using the mean squared Hilbert-Schmidt distance as a 
  figure of merit}
\label{sec:hilbertschmidt}
While we have focussed on the average fidelity and the average Quantum
Chernoff bound as our figures of merit for tomography schemes, there
is one commonly used figure of merit that does behave quite
differently and does affect our results --- the mean squared
Hilbert-Schmidt distance. 

Roy and Scott \cite{Roy2007} found Mutually Unbiased Bases (e.g. the cube
measurement for a single qubit in our terminology) were an optimal
measurement set for single qubit tomography using the mean squared
Hilbert-Schmidt distance as a figure of merit. (They also neglect any
effect of the positivity constraint on the density matrix, but our
simulations bear out the hope that this is negligible.) These
conclusions contrast to ours based on average fidelity which show that
higher order Platonic solids perform better than cube measurements.

To explore this discrepancy we repeated the calculation of
(\ref{eqn:MixedAverageCubeFidelity}) (which also neglects the
positivity constraints) replacing the fidelity with squared
Hilbert-Schmidt distance to give a point squared Hilbert-Schmidt
distance of:
\begin{equation}
\label{eqn:atomictracedist}
\langle D_{hs}^2(r) \rangle \ge \frac{3}{2N}(3 - r^2)
\end{equation}
where $r$ is the radius of the state on the Bloch sphere.  This
equation was derived in \cite{Rehacek2004} by other means and agrees
with the results of
\cite{Roy2007}. Observe the point squared Hilbert-Schmidt distance is
independent of the angular parameters of the state on the Bloch
sphere, unlike the point fidelity case which has an angular
dependence, as shown in Eqn.~(\ref{eqn:MixedAverageCubeFidelity}).

We have repeated our numerical simulations using mean squared
Hilbert-Schmidt distance as a figure of merit, a Bures prior and full
maximum likelihood estimation.  Our results show that for atomic
qubits 
the Platonic solids all
performed equivalently. Indeed our results 
were 
within statistical errors of the theoretical value obtained by 
integrating (\ref{eqn:atomictracedist}) over a Bures prior
$D_{B} \ge 0.8438/N$. This suggests that the result of 
\cite{Roy2007} that the cube measurement is optimal for mean squared
Hilbert-Schmidt distance squared holds, even when the positivity
constraint is included. We also note that for single-detector
configurations our conclusions still hold, the James4 measurement set
still performs badly and should be replaced by a cube measurement.

\section{Discussion and conclusions}
\label{sec:conclusions}

\subsection{Experimental Recommendations}

The clear conclusion of this paper is that for single-detector photonic qubit
tomography, experimentalists should immediately switch from the minimal four
measurement set of \cite{James2001} to the overcomplete six measurement cube
set, or at the very least switch to the tetrahedron measurement set if the
number of measurement settings is an issue.  Moving to the dual-detector
configuration would further improve the results (subject to statistical
fluctuation being the major source of error).

There is also a consistent trend towards better asymptotic performance
for higher order Platonic solids for single qubit states.
Experimentalists would have to judge whether there is a net gain after
time to switch measurement settings is taken into account. We note
that measurements on trapped ions are generally performing what we
call the cube measurement set already.

We have found that this trend is maintained in two-qubit experiments
with tensor product Platonic solid measurements performing very well
even in comparison to entangled measurement sets. This strongly
suggests that the use of these symmetrically arranged measurements
will be advantageous for experiments on systems having larger numbers
of qubits. In this case we note that the number of measurement
settings grows very rapidly and the time to change between detector
settings is likely to become an issue.  In this regard the tetrahedron
measurement should have much improved performance but requires no more
measurements than the generalisations of James4 and James16.

We also note that while experiments in tomography usually report an
uncertainty in the reconstruction based on a fixed number of measured
systems, we are not aware of studies of how this uncertainty is
reduced over time during the tomography. Such a study should contain
information about the noise sources affecting the
experiment. Specifically drifts in either the states or the
measurement over time might be expected to lead to a saturation of the
average fidelity with $N$. We have seen that when the experiment is
limited by statistical noise then the quality of the tomography
depends systematically on $N$ and on the choice of measurement. This
could be tested experimentally and should give detailed information
about the true errors in real experiments. In the case of optical
experiments this simply requires rotating regularly among the
different measurement settings and keeping track of the order in which
counts arrive.  
State inversions can then be done at various time intervals and a
graph of  the number of qubits $N$ against point fidelity similar to our
Fig.~\ref{Fig:AtomicOffAxisPointFidelityVsRadius} could be
straightforwardly obtained.  One should, for example, be able
to observe the general behaviour of a scaling like $1/\sqrt{N}$ at low
numbers of copies transitioning to a scaling of $1/N$ at large numbers
of copies. Likewise the improvement of the point fidelity with the
choice of measurement could be checked.
Comparison to detailed point fidelity simulations should aid
characterization of experimental error sources.

\subsection{Future work}

One could certainly consider measurements made from polyhedra other
than the Platonic solids (for example the semi-regular polyhedra).
However this paper has shown that the Platonic solids provide a good
spread in the number of measurements and performance, and going to
even higher order shapes appears to be of decreasing value. We thus
consider exploring other shapes to be unfruitful with one likely
exception. The SicPOVM was designed to be a spherical $t$ design with
$t=2$.  It is well known \cite{Hardin1996} that the cube and octagon
are 3-designs while the dodecahedron and icosahedron are 5-designs. It
is likely that this is the origin of the similarity in performance of
the dual Platonic solids and $t$ is an indicator of the quality of
tomography performance. This being the case it would be well worth
investigating McLaren's ``improved Snub Cube'' \cite{Hardin1996} which
is a 24 point 7-design.

\section*{Acknowledgements}
We would like to thank Gerard Milburn for useful discussions, and the
Australian Research Council for funding this work.

\appendix
\section{ Derivation of Pure State Cram\'{e}r-Rao bounds for 
  single qubit Tomography }
\label{app:cramerrao}

Let $\eta = [\theta, \phi]^T$ be the state Bloch vector in spherical
polar coordinates, and let $\hat{\eta}$ be the state estimate.  The
similarity between $\eta$ and $\hat{\eta}$ is given by the fidelity
$f_\eta(\hat{\eta}) = (1 + r.\hat{r})/2$ where $r =
[\cos(\phi)\sin(\theta), \sin(\phi)\sin(\theta), cos(\theta)]^T$ and
similarly $\hat{r} = [\cos(\hat{\phi})\sin(\hat{\theta}),
\sin(\hat{\phi})\sin(\hat{\theta}), cos(\hat{\theta})]^T$

Since we assume $\eta$ and $\hat{\eta}$ are close we expand the
fidelity in a Taylor series to second order giving:
\begin{equation}
f_\eta(\hat{\eta}) \approx f_\eta(\eta) + Df_\eta(\eta).(\hat{\eta} - \eta)
+ \frac{1}{2}(\hat{\eta} - \eta)H(\eta)(\hat{\eta} - \eta)^T
\end{equation}
where $Df_\eta(\eta)_i = \frac{\partial f_\eta(\hat{\eta})} { \partial
  \eta_i } |_{\hat{\eta} = \eta}$ and $H(\eta)$ is the Hessian matrix
defined by:
\begin{equation}
\label{Eq:HessianMatrix}
H(\eta)_{i,j} = 
\frac{\partial^2 f_{\eta}(\hat{\eta}) }{\partial \hat{\eta}_i \partial \hat{\eta}_j} 
\big |_{\hat{\eta} = \eta}.
\end{equation}
Now $f_\eta(\eta) = 1$ and $Df_\eta(\eta) = 0$ so

\begin{equation}
f_\eta(\hat{\eta}) \approx 1 + \frac{1}{2} \sum_{i,j} \left . 
\frac{\partial^2 f_\eta(\hat{\eta}) }{\partial \eta_{i} 
\partial \eta_{j}} 
\right  \rvert_{ \hat{\eta} = \eta}
(\hat{\eta}_{i} - \eta_{i})(\hat{\eta}_{j} - \eta_{j}).
\end{equation}

During tomography, each of $N$ identical copies of the qubit state are
each measured exactly once. Our measurement set consists of $l$
different measurement settings. So $\mathcal{N} = N/l$ measurements
are performed with each measurement setting.

Each measurement setting is represented by Bloch vector $\vec{m_i}$,
with outcomes $\chi_i{=}\pm{1}$.  One set of measurements has $2^l$
possible outcomes represented by the string $\chi = \chi_1..\chi_l$.
The probability of obtaining outcome $\chi$ is given by:
\begin{equation}
\label{Eq:outcomProbabilities}
p_{\eta}(\chi) = \prod_i (1 + \chi_i \vec{r}.\vec{m_i})/2.
\end{equation}

The point fidelity obtained by averaging over the possible outcomes
$\chi$ of the set of measurements is given by
\begin{eqnarray}
F(\eta) &=& \sum_{\chi} p_{\eta}(\chi) f_{\eta}(\hat{\eta}(\chi)) \\
                                &\approx& 1 + \frac{1}{2} \mathrm{tr}[H(\eta)V(\eta)] 
\end{eqnarray}
where $V(\eta)$ is the covariance matrix defined by:
\begin{equation} 
\label{Eq:CovarianceMatrix}
V(\eta) = \sum_\chi p(\chi) (\hat{\eta}(\chi) - \eta) (\hat{\eta}(\chi) - \eta)^T.
\end{equation}
 
The Cram\'{e}r-Rao bound for unbiased estimators states:
\begin{equation}
\label{Eq:CramerRaoBound}
V(\eta) \ge  \frac{1}{ \mathcal{N} I(\eta)}
\end{equation}
where $I(\eta)$ is the Fisher Information matrix defined by:
\begin{equation}
\label{Eq:FisherInformation}
I_{ij}(\eta) = \sum_\chi p_{\eta}(\chi)\frac{ \partial \ln{p_{\eta}(\chi)}}{ \partial \eta_i}
\frac{ \partial \ln{p_{\eta}(\chi)}}{ \partial \eta_j}.
\end{equation}

Noting that $H(\eta)$ is negative definite, the point fidelity bound
is then given by:
\begin{equation}
F(\eta) \le 1 + \frac{1}{2\mathcal{N}} \mathrm{tr}[H(\eta)I^{-1}(\eta)]. 
\end{equation}

Finally averaging over $\eta$ uniformly according to the Haar measure:
\begin{equation}
d\rho = \frac{\sin{\theta}}{4\pi} d\theta d\phi
\end{equation}
gives average fidelity Cram\'{e}r-Rao Bound.

\section{ Eigenvalue distribution for the quantum Chernoff bound metric }
\label{sec:chenoffevals}
In \cite{Audenaert2006} it was shown that the infinitesimal distance
between states $\rho$ and $\rho{+}d\rho$ according to the Quantum
Chernoff bound is:
\begin{equation}
\label{Eqn:QuantumChernoffBoundMetric}
ds^2 = \frac{1}{2}\sum_{jk} \frac{|\bra{j} d\rho \ket{k}|^2}
{(\sqrt{\lambda_j} + \sqrt{\lambda_k})^2}
\end{equation}
where $\rho = \sum_j \lambda_j \ket{j} \bra{j}$ is the eigenvalue
decomposition of $\rho$. We will follow the same procedure as
\cite{Hall1998} in his calculation of the Bures volume element to
derive our quantum Chernoff bound eigenvalue distribution.
 
First decompose this into an infinitesimal shift in eigenvalues
followed by an infinitesimal unitary rotation:
\begin{eqnarray}
\rho + \delta\rho &=& (I + \delta U)(\rho + \delta \Lambda)(I + \delta U)^\dag \\
&=& \rho + \delta \Lambda + [\delta U, \rho] 
\label{Eqn:InfinitesimalDensityShift}
\end{eqnarray}
where we have ignored any terms which are a product of multiple
infinitesimal quantities and defined $\bra{j} \delta \Lambda \ket{k} =
\delta_{jk} d\lambda_j$.  In general we can rewrite the infinitesimal
unitary operator as:
\begin{equation}
\label{Eqn:InfinitesimalUnitary}
\delta U = \sum_{j\le k}(dx_{jk} + i dy_{jk})\ket{j}\bra{k} + h. c. 
\end{equation}
Next we substitute Eq.~\ref{Eqn:InfinitesimalUnitary} and
Eq.~\ref{Eqn:InfinitesimalDensityShift} into
Eq.~\ref{Eqn:QuantumChernoffBoundMetric} yielding
\begin{eqnarray}
ds^2 &=& \frac{1}{8}\sum_j \frac{(d\lambda_j)^2}{\lambda_j}
+  \sum_{j < k}\frac{(\lambda_k - \lambda_j)^2}
{(\sqrt{\lambda_j} + \sqrt{\lambda_k})^2}(dx_{jk}^2 + dy_{jk}^2). \nonumber
\end{eqnarray}
We convert this into a differential volume element by multiplying
together all the $ds_i$ formed by a shift in each parameter
$\lambda_i$ assuming the shift in the other parameters is zero. Since
the volume element will require normalization we combine any constant
factors. The final differential volume element is given by:
\begin{equation}
dV = C^\prime \frac{d\lambda_1..d\lambda_M}{\sqrt{\lambda_1..\lambda_M}}
\prod_{k<j} \frac{(\lambda_k - \lambda_j)^2}{(\sqrt{\lambda_j} + \sqrt{\lambda_k})^2}
dx_{jk} dy_{jk}
\end{equation}
where $C^\prime$ is a normalisation constant.

Since the eigenvalue distribution is unitarily invariant we can
concern ourselves with the marginal probability distribution over the
eigenvalues of the density operators. That is:
\begin{equation}
p(\lambda_1..\lambda_M) = \frac{C}{\sqrt{\lambda_1..\lambda_M}}
\prod_{k<j} \frac{(\lambda_k - \lambda_j)^2}{(\sqrt{\lambda_j} + \sqrt{\lambda_k})^2}
\end{equation}
where the eigenvalues are also constrained to satisfy $\sum_{j=1}^M
\lambda_j = 1$ and $C$ is a normalisation constant. Next we examine the
single qubit case.

\subsection{Single Qubit Case}

For the qubit case we have:
\begin{equation}
p(\lambda_1\lambda_2) = \frac{C}{\sqrt{\lambda_1\lambda_2}}
\frac{(\lambda_1 - \lambda_2)^2}{(\sqrt{\lambda_1} + \sqrt{\lambda_2})^2}
\end{equation}

The eigenvalues for the qubit case are related to $r$ the radius on the
Bloch sphere according to: $\lambda_1{=}(1{+}r)/2$ and
$\lambda_2{=}(1{-}r)/2$. This yields

\begin{equation}
p(r) = \frac{C}{2}\frac{1-\sqrt{1-r^2}}{\sqrt{1-r^2}}   
\end{equation}

By requiring $\int_0^1 p(r) dr = 1$ we calculate $C{=}4/(\pi{-}2)$.

\end{document}